\documentclass[a4paper,dvips,12pt]{article}
\usepackage{amsmath,latexsym,amssymb,epsfig,psfrag}  
\usepackage{comment}
\usepackage{cite}
\usepackage{wasysym}

\newcommand{\be}{\begin{equation}}
\newcommand{\ee}{\end{equation}}

\newcommand{\de}{\partial}

\begin{document}  
  
\baselineskip=16pt   
\begin{titlepage}  
\begin{center}  
\hfill{{\bf UAB-FT-559}}

\hfill{{\bf IPPP/04/14}}

\hfill{{\bf DCTP/04/28}}

\hfill{{\bf DESY-04-059}}

\vspace{0.5cm}  
  
\large {\sc \Large Fermions and Supersymmetry Breaking  \\
\Large  in the Interval}

\vspace*{5mm}  
\normalsize  
  
{\bf G.~v.~Gersdorff~\footnote{g.v.gersdorff@desy.de}, 
L.~Pilo~\footnote{luigi.pilo@pd.infn.it},
M.~Quir\'os~\footnote{quiros@ifae.es}, 
A.~Riotto~\footnote{antonio.riotto@pd.infn.it},
V.~Sanz~\footnote{veronica.sanz-gonzalez@durham.ac.uk}}   

\smallskip   
\medskip   
\it{~$^{1}$~Deutsches Elektronen-Synchrotron, Notkestrasse 85}\\ 
\it{D-22603 (Hamburg) Germany}

\smallskip    
\medskip  
\it{~$^{2,\,4}$~Department of Physics and INFN, Sezione di Padova \\
via Marzolo 8, I-35131 (Padova) Italy}

\smallskip   
\medskip   
\it{~$^{3}$~Instituci\'o Catalana de Recerca i Estudis Avan\c{c}ats (ICREA)}\\
\it{and}\\
\it{Theoretical Physics Group, IFAE/UAB}\\ 
\it{E-08193 Bellaterra (Barcelona) Spain}

\smallskip   
\medskip   
\it{~$^{5}$~}IPPP Physics Department, Durham University\\ 
\it{DH1 3LE (Durham) UK}

\vskip0.6in \end{center}  
   
\centerline{\large\bf Abstract}  

\noindent
We study fermions, such as gravitinos and gauginos in supersymmetric
theories, propagating in a five-dimensional bulk where the fifth
dimensional component is assumed to be an interval. We show that the
most general boundary condition at each endpoint of the interval is
encoded in a single complex parameter representing a point in the
Riemann sphere. Upon introducing a boundary mass term, the variational
principle uniquely determines the boundary conditions and the bulk
equations of motion. We show the mass spectrum becomes independent
from the Scherk-Schwarz parameter for a suitable choice of one of the
two boundary conditions. Furthermore, for any value of the
Scherk-Schwarz parameter, a zero-mode is present in the mass spectrum
and supersymmetry is recovered if the two complex parameters are
tuned.

\vspace*{2mm}

\end{titlepage}  
  

\noindent
A common feature of five-dimensional supersymmetric models are
fermions propagating in the bulk of the extra dimension.  In order to
extract physical predictions at low energies, the four dimensional
mass spectrum of those fermions has to be known.  For instance,
supersymmetry breaking is determined by the mass spectrum of the
gravitino, the existence of a zero mode signalling unbroken
supersymmetry. Similarly, when gauge multiplets propagate in the bulk
supersymmetry breaking is intimately linked to the existence of
gaugino zero modes. In particular if supersymmetry breaking is
implemented by non-trivial twist conditions, or Scherk-Schwarz
mechanism~\cite{ss}, it acts in the same way both in the gravitino and
the gaugino sectors.

The aim of this letter is to study fermions propagating in a flat
five-dimensional space-time, with coordinates $(x^\mu,y)$, where the
compact fifth dimension (with radius $R$) has two four-dimensional
boundaries located at $y=0$ and $y=\pi R$.  Often this space is
constructed as the orbifold $S^1/\mathbb Z_2$, identifying points on
the circle related by the reflection of the fifth coordinate $y\to
-y$. Fields with odd parity with respect to the $\mathbb Z_2$
reflections are zero at the fixed points, while the normal derivative
of even fields is forced to vanish.  The treatment of fermions is
complicated in the presence of brane actions localized at the
boundaries.  In the orbifold approach, these brane actions are
introduced with a delta-function distribution, peaked at the location
of the orbifold fixed point.  The latter induces discontinuities in the
wave functions of the fermions which take different values at the
fixed point and infinitesimally close to
it~\cite{Bagger:2001qi,Delgado:2002xf}. A possible way to avoid these
jumps is to give up the rigid orbifold boundary conditions
and instead enforce the fields to be continuous, while the boundary
conditions are determined by the boundary action itself.  This is
called the {\em interval} approach and leads to physically equivalent
spectra as those of the orbifold approach without any need of using,
as the latter, singular functions~\footnote{The interval approach is
sometimes called ``downstairs'' approach while the orbifold approach
is called ``upstairs'' approach.}.  To summarize, in the orbifold
approach one imposes fixed (orbifold) boundary conditions while the
brane action induces jumps, whereas in the interval approach
one imposes continuity and the brane action induces the boundary
conditions.

In this letter we will follow the interval approach and show how the
boundary action can give rise to consistent boundary conditions for
the fermions. In a forthcoming publication~\cite{gpqrs2} we will give
a detailed treatment of how to translate the two pictures into each
other.  In a manifold ${\cal M}$ with a boundary the dynamics is
determined by two equally important ingredients: the bulk equations of
motion and the boundary conditions (BC's). An economical way to
determine a set of consistent BC's together with the bulk equations of
motion is the action principle~\footnote{For an alternative approach
see~\cite{csfermion}.}: under a variation of the dynamical fields the
action must be stationary. This in general translates into two
separate conditions: the vanishing of the variation of the action in
the bulk and the vanishing of the variation at the boundary $\partial
{\cal M}$. Contributions to the action variation at the boundary come
from integration by parts of bulk variation and, if present, from
varying the boundary part of the action (see~\cite{Csaki:2003dt} for a
recent application to symmetry breaking).  In the following we will
consider the five-dimensional (5D) manifold ${\cal M}$ as the direct
product of the four dimensional Minkowski space ${\cal M}_4$ and the
interval $[0, \pi R]$.

Since we are mainly interested in supersymmetric theories, we will
take the fermions to be symplectic-Majorana spinors, although a very
similar treatment holds for the case of fermionic matter field
associated to Dirac fermions.  In particular we will consider the
gaugino case, the treatment of gravitinos being completely analogous.
The 5D spinors $\Psi^i$ satisfy the symplectic-Majorana reality
condition and we can represent them in terms of two chiral 4D spinors
according to~\footnote{We use the Wess-Bagger convention~\cite{wb} for
the contraction of spinor indices.}
\begin{equation}\label{uno}
\Psi^i = \begin{pmatrix} \eta^i_\alpha \\ 
\bar{\chi}^{i\,\dot{\alpha}} \end{pmatrix} \, , \qquad
\bar{\chi}^{i \,\dot{\alpha}} \equiv \epsilon^{ij} \, 
\left(\eta^j_\beta \right)^\ast \,  \epsilon^{\dot{\alpha} 
\dot{\beta}}  
\; .
\end{equation}
where $\epsilon_{ij}= i \,(\sigma_2)_{ij}$ and $\epsilon^{im}
\epsilon_{jm} = \delta^i_j$.  Consider thus the bulk Lagrangian
\begin{equation} 
{\cal L}_{\text{bulk}} =i\, \bar{\Psi} \gamma^{{}_M}
D_{{}_M} \Psi =\frac{i}{2} \bar{\Psi} \gamma^{{}_M}
D_{{}_M} \Psi - \frac{i}{2} D_{{}_M} \bar{\Psi} \gamma^{{}_M} \Psi \; .
\end{equation}
where the last equation is not due to partial integration but holds
because of the symplectic-Majorana property, Eq.~(\ref{uno}). The
derivative is covariant with respect to the $SU(2)_R$ automorphism
symmetry and thus contains the auxiliary gauge connection $V_M$.  The
field $V_M$ is non propagating and appears in the off-shell
formulation of 5D supergravity~\cite{zucker}.  A vacuum expectation
value (VEV)~\footnote{Consistent with the bulk equation of motion
  $d\,(\vec q\cdot \vec V)=0$~\cite{zucker}.}
\be V_{_M} = \delta^5_{_M} \, \frac{\omega}{R} \,
\vec{q}\cdot\vec\sigma \, , \qquad {\vec{q}}^{\, 2}=1
\label{SSVEV}
\ee
implements a Scherk-Schwarz (SS) supersymmetry breaking
mechanism~\cite{ss} in the Hosotani basis~\cite{hos,sshos}. The
standard form of the SS mechanism, originally introduced for circle
compactification, can be recovered by a gauge transformation $U$ that
transforms away $V_{_M}$ but twists the periodicity condition for
fields charged under $SU(2)_R$ on the circle.  As we will see later in
the interval a SS breaking term is equivalent to a suitable
modification of the BC's at one of the endpoints.  The unitary vector
$\vec{q}$ points toward the direction of SS breaking.  We supplement
the bulk action by the following boundary terms at $y=y_f$ ($f=0,\pi$)
with $y_0=0$ and $y_\pi=\pi R$
\begin{equation} 
{\cal L}_{f} = \frac{1}{2} \bar{\Psi} \left(T^{{(f)}} + \gamma^5 \,
V^{{(f)}} \right) \Psi =\frac{1}{2} \,{\eta^i}M^{_{(f)}}_{ij} \eta^j
+{\rm h.c.}\; , 
\label{bmasses}
\end{equation}
where $T^{{(f)}}$ and $V^{{(f)}}$ are matrices acting on $SU(2)$
indices,
\begin{equation}
M^{{(f)}}=i\sigma_2 \, (T^{{(f)}}-i V^{{(f)}})
\label{TVm}
\end{equation}
and we have made use of the decomposition (\ref{uno}).  Notice that
the mass matrix is allowed to have complex entries.  Without loss of
generality we take it to be symmetric, which enforces $T^f$ and $V^f$
to be spanned by Pauli matrices.

The total bulk + boundary action is then given by
\begin{equation}
S= S_{\text{bulk}}+S_{\text{boundary}}=\int d^5x \, {\cal
L}_{\text{bulk}} + \int_{y =0} d^4x \,{\cal L}_{0} - \int_{y =\pi R}
d^4x \,{\cal L}_{\pi} \quad .
\label{total}
\end{equation} 
The variation of the bulk action gives 
\begin{equation}
\delta S_{\text{bulk}} = \int d^5x \, i \left(\delta \bar{\Psi}
\gamma^{{}_M} D_{{}_M} \Psi - D_{{}_M} \bar{\Psi} \gamma^{{}_M}
\delta\Psi \right) - \int d^4x \, \left[\delta \eta^i \,\epsilon_{ij} \eta^j 
 + h.c. \right]^{\pi R}_0 \; ,
\label{bvar}
\end{equation}
where the boundary piece comes from partial integration.  One now has
to add the variation of the boundary action.  Enforcing that the total
action $S= S_{\text{bulk}}+ S_{\text{boundary}}$ has zero variation we
get the standard Dirac equation in the bulk provided that all the
boundary pieces vanish. The latter are given by
\begin{equation}
\left.\left[\delta \eta^i \left( \epsilon_{ij} +
M_{ij}^{{}_{(f)}}\right) \eta^j + \, \text{h.c.}  \right]\right|_{y
= y_f} = 0 \; . \label{bf}
\end{equation}
Since we are considering unconstrained variations of the fields, the
BC's we obtain from Eqs.~(\ref{bf}) are given by
\begin{equation}
\left.\left( \epsilon_{ij} + M_{ij}^{{}_{(f)}}\right) \eta^j
\right|_{y = y_f} = 0 \; . \label{bf2}
\end{equation}
These equations only have trivial solutions (are overconstrained)
unless
\be
\det\left(\epsilon_{ij} + M_{ij}^{{}_{(0)}}
 \right) =\det\left( \epsilon_{ij} + M_{ij}^{{}_{(\pi)}}\right)=0\;.
\label{det}
\ee
Imposing these conditions, we get the two complex BC's which are
needed for a system of two first order equations.  Note that this
means that an arbitrary brane mass matrix does not yield viable BC's;
in particular a vanishing brane action is inconsistent~\footnote{In
the sense that the action principle does not provide a consistent set
of BC's as boundary equations of motion.} since
$\det(\epsilon_{ij})\neq 0$~\footnote{Notice that this agrees with the
methods recently used in Ref.~\cite{moss}.}.  However this does not
imply that the familiar orbifold BC $\eta_1=0$ ($\eta_2=0$) can not be
achieved; in the interval approach they correspond to $M=\sigma^1$
($M=-\sigma^1$).

The BC's resulting from Eqs.~(\ref{bf2}) are of the form
\begin{equation}
\left.\left(c^1_f \, \eta^1 + c^2_f \, \eta^2 \right) \right|_{y =y_f}
= 0 \quad ,
\label{cbc}
\end{equation}
where $c^{1,2}_f$ are complex parameters or, setting $z_f =- (c^1_f/c^2_f)$
\begin{equation}
\left.\left(\eta^2 -z_f \, \eta^1 \right) \right|_{y =y_f} = 0, \quad z_f\in
\mathbb C\quad .
\label{bcf}
\end{equation}
Physically inequivalent BC's span a complex projective space
$\mathbb{C}P^1$ homeomorphic to the Riemann sphere.  In particular,
$z_f =0$ leads to a Dirichlet BC for $\eta_2$, and the point at
infinity $ z_f =\infty$ leads to a Dirichlet BC for $\eta_1$.  Notice
that these BC's come from $SU(2)_R$ breaking mass terms. Special
values of $z_f$ correspond to cases when these terms preserve part of
the symmetry of the original bulk Lagrangian. In particular when both
the SS and the preserved symmetry are aligned those cases can lead to
a {\it persistent} supersymmetry as we will see.  Once (\ref{det}) is
satisfied, the values of $z_f$ in terms of the brane mass terms are
given by
\be 
z_f=-\frac{M_{11}^{_{(f)}}}{1+M_{12}^{_{(f)}}}
 =\frac{1-M_{12}^{_{(f)}}}{M_{22}^{_{(f)}}}
\label{zBC}
\ee
where the second equality holds due to the condition (\ref{det}).

The mass spectrum is found by solving the EOM with the boundary
conditions (\ref{bcf}).  To simplify the bulk
equations of motion it is convenient to go from the Hosotani basis
$\Psi^i$ to the SS one $\Phi^i$, related by the transformation
\begin{equation} 
\Psi = U \, \Phi ,\quad U=
\exp{\left(-i\, \vec{q} \cdot \vec{\sigma}\, \omega\, \frac{y}{R}\right)}  \;. 
\label{cb}
\end{equation}
In the SS gauge the bulk equations read 
\begin{equation} 
i \, \gamma^{{}_M} \de_{{}_M} \Phi = 0 \quad .
\end{equation}
We now decompose the chiral spinor $\eta^i(x,y)$ in the Hosotani basis
as $\eta^i(x,y) = \varphi^i(y) \psi(x)$, with $\psi(x)$ a 4D chiral
spinor. Setting $\varphi = U \phi$ we get the following equations of
motion in the SS basis
 \begin{equation}
 m \,  \phi^i -   \epsilon^{ij}  \, \frac{d \bar{\phi}_j}{dy} = 0 \, , \quad  
 m \,  \bar{\phi}_j \, \epsilon^{ij} + \frac{d \phi^i}{dy} = 0 \; . 
\label{bulkeom}
 \end{equation}
The parameter $m$ in Eq.~(\ref{bulkeom}) is the Majorana
mass eigenvalue of the 4D chiral spinor~\footnote{The bar acting on a
scalar quantity, as~e.g.~$\bar\phi_i$, and a chiral spinor,
as~e.g.~$\bar\psi$, denotes complex conjugation.}
\begin{equation}
i \sigma^\mu \de_\mu \bar{\psi} = m \, \psi \, , \qquad i
\bar{\sigma}^\mu \de_\mu \psi = m \, \bar{\psi} \quad .
\end{equation}
As a consequence of the transformation (\ref{cb}) the SS parameter
$\omega$ manifests itself only in the BC at $y=\pi R$~\footnote{Notice
that $U(y=0)=1$.  The roles of the branes and hence of $z_\pi $ and
$z_0$ can be interchanged by considering the SS transformation
$U'(y)\equiv U(y-\pi R)$.}:
\be \zeta_0\equiv\left.\frac{\phi^2}{\phi^1}\right|_{y=0}=z_0,\quad
\zeta_\pi\equiv\left.\frac{\phi^2}{\phi^1}\right|_{y=\pi R}=\frac{
\tan(\pi\omega) (i q_1 - q_2 - i q_3\, z_\pi) + z_\pi } {\tan(\pi\omega) (i q_1
\, z_\pi + q_2\, z_\pi + i q_3) + 1 }\;,
\label{SSbc}
\ee 
where $\zeta_f$ are the BC's in the SS basis. In particular the
boundary condition $\zeta_\pi$ is a function of $\omega$, $\vec q$ and
$z_\pi$.  From this it follows that we can always gauge away the SS
parameter $\omega$ in the bulk Lagrangian going into the SS basis
through (\ref{cb}). However now in the new basis $\omega$ reappears in
one of the BC's.

The bulk equations have the following generic solution
\begin{equation}
 \phi(y)= 
  \begin{pmatrix}     
    \bar a \cos(my)+\bar z_0 a \sin(my)
    \\
    -a\sin(my)+z_0 \bar a \cos(my)
  \end{pmatrix} \quad ,
 \label{sol}
\end{equation}
where $a$ is a complex number given in terms of $z_0$ and $\zeta_\pi$:
\begin{equation}
  a=\frac{z_0-\zeta_\pi}{|z_0-\zeta_\pi|}+\frac{1+z_0 \bar
  \zeta_\pi}{|1+z_0\bar \zeta_\pi |}\; .
\end{equation}
The solution (\ref{sol}) satisfies the BC's Eq.~(\ref{SSbc}) for the
following mass eigenvalues
\begin{equation}
m_n  = \frac{n}{R} + \frac{1}{ \pi
R }\,\arctan\left|\frac{z_0-\zeta_\pi}{1+z_0\,\bar \zeta_\pi}\right|\; ,
\label{masses}
\end{equation}
where $n\in \mathbb Z$. When $z_0=\zeta_\pi$ there is a zero mode and
supersymmetry remains unbroken. When the only sources of supersymmetry breaking
reside on the branes, setting them to cancel each other, $z_0=z_{\pi}$,
preserves  supersymmetry~\cite{horava}. Once supersymmetry is further 
broken in the bulk, an obvious way to restore it is by
determining $z_\pi$ as a function of $z_0$ and $\omega$ using the
relation (\ref{SSbc}) with $\zeta_\pi=z_0$.  This will lead to an
$\omega$-dependent brane-Lagrangian at $y=\pi R$. In this case we
could say that supersymmetry, that was broken by BC's (SS twist) is
{\it restored} by the given SS twist (BC's)~\cite{SSBC}.

There is however a more interesting case: suppose the brane Lagrangian
determines $z_\pi$ to be
\begin{equation}
z_\pi=z(\vec q\,)\equiv\frac{\lambda-  q_3}{q_1 - i q_2}
\label{vH}.
\end{equation}
with $\lambda=\pm 1$.  This special value of $z_\pi$ is a fixed point
of the SS transformation, i.e.~$\zeta_f=z_f$.  For $z_\pi=z(\vec q\,)$
the spectrum becomes independent on $\omega$.  In other words, for
this special subset of boundary Lagrangians, the VEV for the field
$\vec q\cdot\vec V_5$ does not influence the spectrum. The reason for
this can be understood by going back to the Lagrangian which we used
to derive the BC's. From the relation (\ref{zBC}) one can see that
condition (\ref{vH}) is satisfied by the mass matrix
\begin{eqnarray}
M_{12}^{_{(\pi)}} &=& \lambda q_3 \nonumber \\
M_{11}^{_{(\pi)}} &=& -\lambda (q_1 + i q_2 ) \nonumber \\
M_{22}^{_{(\pi)}} &=& \lambda(q_1 -  i q_2 ) 
\end{eqnarray} 
which can be translated into a mass term at the boundary $y=y_\pi$
along the direction of the SS term, i.e.~$V^{(\pi)}=0$ and $T^{(\pi)}=
- \lambda\,\vec q\cdot\vec\sigma$ in the notation of
Eq.~(\ref{bmasses}). In particular this brane mass term preserves a
residual $U(1)_R$ aligned along the SS direction $\vec{q}$.  In other
words, the SS-transformation $U$ leaves both brane Lagrangians
invariant and $\omega$ can be gauged away.  When we further impose
$z_0=z(\pm\vec q\,)$, i.e.~$V^{(0)}=0$ and $T^{(0)}=\pm T^{(\pi)}$ the
$U(1)_R$ symmetry is preserved by the bulk. In particular if
$z_0=z(\vec q\,)$ supersymmetry remains unbroken, although the VEV of
$\vec q\cdot \vec V_5$ is nonzero. One could say that in this case the
theory is {\it persistently} supersymmetric even in the presence of
the SS twist, with mass spectrum $m_n=n/R$. On the other hand if
$z_0=z(-\vec q\,)$ the theory is ({\it persistently})
non-supersymmetric and independent on the SS twist: the mass spectrum
is given by $m_n=(n+1/2)/R$. In this case supersymmetry breaking
amounts to an extra $\mathbb Z_2^\prime$
orbifolding~\cite{Barbieri:2000vh}.

Notice that we have not chosen the most general solution to
Eq.~(\ref{vH}) but one where $V^{(f)}=0$. In the most general case the
condition~(\ref{det}) leads to $(\vec{T}^{(f)})^2-(\vec{V}^{(f)})^2=1$
and $\vec{T}^{(f)}\cdot\vec{V}^{(f)}=0$, and for $\vec{V}^{(f)}\neq 0$
Eq.~(\ref{vH}) has in general a two-parameter family of solutions. All
of them should comply with the existence of {\it persistent} zero
modes (irrespective of the SS twist). However the condition for an
(off-shell) supersymmetric action is only consistent with the solution
with $V^{(f)}=0$, as we will see below.

Something similar happens in the warped case~\cite{gpqrs2}: when bulk
cosmological constant and brane tensions are turned on, invariance of
the action under local supersymmetry requires gravitino mass terms on
the brane. In the tuned case, -- i.e.~in the Randall-Sundrum (RS)
model -- those brane mass terms precisely give rise to the BC
$z_0=z_\pi=z(\vec q)$~\cite{bagger1}.  Note that there $\vec
q\cdot\vec V_5$ is replaced by $A_5$, the fifth component of the
graviphoton.  In fact, it has been shown that in this case there
always exists a Killing spinor and supersymmetry remains
unbroken~\cite{bagger2,lalak}, consistent with the result that in RS
supersymmetry can not be spontaneously broken~\footnote{A discrete
supersymmetry breaking by BC's, $z_0=z(-\vec q)$, $z_\pi=z(\vec q\,)$,
was performed in Ref.~\cite{Gherghetta:2000kr}.}  by the SS
mechanism~\cite{Hall:2003yc,bagger1}.  This and other issues, such as
the comparison between the interval and the orbifold approaches and
how to relate them, will be presented elsewhere~\cite{gpqrs2}.

Up to now, we have focused on the fermion sector spectrum. Adding the
complete vector multiplet does not invalidate our conditions for
supersymmetry restoration as long as the supersymmetry breaking brane
terms are of the form given by Eq.~(\ref{bmasses}). We would like to
show the invariance of our gaugino Lagrangian, Eq.~(\ref{total}),
under (global) supersymmetry.  To this end, let us focus on a simple
abelian gauge multiplet. Clearly, since we are not imposing any a
priori boundary condition on the fields in the action, we have to
worry about the total derivatives which arise in the variation of the
bulk action. The latter is given by~\footnote{Besides the gauge field
$B_M$ with field strength $G_{MN}$ and the gaugino $\Psi$ the 5D
vector multiplet contains the real scalar $\Sigma$ and the auxiliary
$SU(2)_R$ triplet $\vec X$.}
\be S_{\rm bulk}^{U(1)}=\int_{\mathcal M}\left(2\vec X\cdot\vec
X-\Sigma\partial^2\Sigma-\frac{1}{2}\partial_M\Sigma\partial^M\Sigma
+i\bar\Psi/\!\!\!\partial\Psi-\frac{1}{4}G_{MN}G^{MN}\right).
\label{globalbulk}
\ee
Under a global supersymmetric transformation the Lagrangian transforms
into a total derivative giving rise to the supersymmetry
boundary-variation:
\be 
\delta_\epsilon S_{\rm bulk}^{U(1)}=\int_{\cal\partial M}\bar
\epsilon i\gamma^5\rho ,\qquad
%
%
%
\rho=\left(i\vec X \cdot \vec\sigma -\Sigma
/\!\!\!\partial-\frac{1}{4}\gamma^{MN}G_{MN}-\frac{1}{2} /\!\!\!
\partial\Sigma\right)\Psi .  
\ee
To compensate for this, we add to it the brane action
\be 
S_{\rm brane}^{U(1)}=\int_{\cal\partial M}\left( 2 \vec T^{(f)}
\cdot\Sigma\vec X+\frac{1}{2}\bar\Psi T^{(f)}\Psi\right) 
\label{globalbrane}
\ee
which transforms into
\be 
\delta_\epsilon S_{\rm brane}^{U(1)}=\int_{\cal\partial
  M}\bar \epsilon\, T^{(f)}\rho .
\ee
Now the supersymmetry variation at each boundary is proportional to
$(1+i\gamma^5 T^{(f)}))\epsilon(y_f)$.  Denoting with $\xi$ [see
Eq.~(\ref{uno})] the upper part of $\epsilon$, whenever $(\vec
T^{(f)})^2=1$ these variations can cancel provided the transformation
parameter satisfies the BC's $\xi^2=z(\vec T^{(f)})\,\xi^1$.  The only
possibility is that $T^{(0)}=T^{(\pi)}$, since $\epsilon$ is constant
for global supersymmetry.  Notice that according to Eqs.~(\ref{det})
and (\ref{TVm}), this gives rise to the same BC's for the gaugino,
$\eta^2=z(\vec T^{(f)})\,\eta^1$.  The remaining EOM then fix the
boundary conditions $G_{\mu 5}=\vec X=\Sigma=0$. The bottom line of
the off-shell approach is that, in the presence of a boundary, at most
one supersymmetry can be preserved.  Global SUSY invariance for the
action of a vector multiplet singles out a special boundary mass term
for gauginos such that $z_0 = z_\pi$ which is at origin of the zero
mode in the spectrum [see Eq.~(\ref{masses}) for $\omega
=0$~\footnote{In the global theory on the interval, all supersymmetry
breaking is encoded in the $T^{(f)}$: there is no auxiliary field
$V_M$ whose VEV could contribute to the breaking .}.]  We expect there
to be a locally supersymmetric extension of the action
(\ref{globalbulk})+(\ref{globalbrane}) for $T^{(0)}\neq T^{(\pi)}$.
In this case the $SU(2)_R$ auxiliary gauge connection $\vec V_M$ from
the supergravity multiplet gives an additional source of supersymmetry
breaking.  Notice that for a globally supersymmetric vacuum there must
then be a solution to the Killing spinor equation
\be
\gamma^5D_5 \epsilon(y)=0,\qquad \xi^2(y_f)=z(\vec
T^{(f)})\,\xi^1(y_f), 
\ee 
where $D_5$ is covariant with respect to $SU(2)_R$. These equations
coincide with the zero mode condition for the gaugino considered
above.

In conclusion we have studied in this letter the issues of fermion
mass spectrum, and supersymmetry breaking in the presence of
Scherk-Schwarz twists~\footnote{We have studied SS or Hosotani
breaking in the bulk, but one could similarly consider radion $F$-term
breaking~\cite{radion}.}, in the interval approach with arbitrary BC's
fixed by boundary mass terms. If {\it alignment\;} occurs, i.e.  BC's
are invariant under the SS twist, the mass spectrum (supersymmetric or
not) becomes independent on the SS parameter. If the BC's are
identical for the different boundaries there appears a zero mode in
the spectrum: supersymmetry is {\it restored} by a cancellation
between BC's and the SS twist. When the two previous conditions are
fulfilled, i.e.~the BC's are equal at different boundaries and SS
twist invariant, the mass spectrum is supersymmetric and independent
on the SS parameter: supersymmetry is {\it persistent} in the presence
of the SS twist. In this case the bulk + brane Lagrangian is invariant
under a remaining $U(1)_R$ symmetry. The conditions imposed on the
brane Lagrangians in the {\it persistent} supersymmetry case can be
regarded as technically natural, since once they are satisfied at tree
level, they will not be upset by corrections coming from the bulk +
brane Lagrangian to any order. Only after the addition of extra
breaking terms, for example brane kinetic terms, supersymmetry would
be broken in a controllable way. Those two conditions could have their
origin on a higher dimensional completion of the theory, as it takes
place at Horava's gaugino condensation model~\cite{horava}, and they
would lead to {\it persistent} supersymmetry after compactification
down to five dimensions.  In our scenario, {\it alignment} would give
rise to a model where supersymmetry could be broken, but the breaking
scale would be completely fixed by the compactification scale $1/R$
and the relative size of brane breaking terms $z_f$, irrespective of
the SS-breaking scale $\omega$. This phenomenon opens new
possibilities for model building whenever one needs to control the
effect of supersymmetry breaking in the bulk.

\vspace*{7mm}
\subsection*{\sc Acknowledgments}

\noindent This work was supported in part by the RTN European Programs
HPRN-CT-2000-00148 and HPRN-CT-2000-00152, and by CICYT, Spain, under
contracts FPA 2001-1806 and FPA 2002-00748 and grant number INFN04-02.
One of us (V.S.) thanks T.~Okui for useful discussions. Three of us
(L.P., A.R. and V.S.) would like to thank the Theory Department of
IFAE, where part of this work has been done, for hospitality.


\end{document}